# Dynamics of desynchronized episodes
# in intermittent synchronization

Leonid L Rubchinsky[1,2], Sungwoo Ahn[1], and Choongseok Park[3]

[1] Department of Mathematical Sciences, Indiana University–Purdue University Indianapolis, Indianapolis, IN 46202

[2] Stark Neurosciences Research Institute, Indiana University School of Medicine, Indianapolis, IN 46202

[3] Department of Mathematics, North Carolina A&T State University, Greensboro, NC 27411

Corresponding author: Leonid L Rubchinsky

Department of Mathematical Sciences, Indiana University–Purdue University Indianapolis

402 N. Blackford St., Indianapolis, IN 46202, USA.

Email: leo@math.iupui.edu

Phone (317) 274-9745

Fax (317) 274-3460

Keywords: synchronization, phase locking, desynchronization, phase space, intermittency

Running title: Desynchronized episodes in intermittent synchronization

Word counts: 2858 (the main body only)

**Abstract**

Intermittent synchronization is observed in a variety of different experimental settings in physics and beyond and is an established research topic in nonlinear dynamics. When coupled oscillators exhibit relatively weak, intermittent synchrony, the trajectory in the phase space spends a substantial fraction of time away from a vicinity of a synchronized state. Thus to describe and understand the observed dynamics one may consider both synchronized episodes and desynchronized episodes (the episodes when oscillators are not synchronous). This mini-review discusses recent developments in this area. We explain how one can consider variation in synchrony on the very short time-scales, provided that there is some degree of overall synchrony. We show how to implement this approach in the case of intermittent phase locking, review several recent examples of the application of these ideas to experimental data and modeling systems, and discuss when and why these methods may be useful.

**Introduction**

Synchronization is observed in a variety of physical phenomena and beyond (see, for example, a book (1)). Examples span from coupled pendula, coupled lasers, and Josephson junctions in physics (1) to various synchronized phenomena in living systems (2), in particular, in neuroscience (3). The latter are also frequently studied by physicists using techniques and methods of physics (4,5).

Synchronization can be defined and measured in different ways (e.g., (1)). In general, the degree of synchrony is higher if the coupling between oscillators is stronger so that eventually a synchronization threshold may be reached. For subthreshold values of coupling (if there is any threshold at all), the oscillators may exhibit intermittent synchronization phenomena, where dynamics is synchronous on some time-intervals and not synchronous on others. This partial, intermittent synchrony may be especially important for biological applications where it may potentially facilitate high adaptability of biological systems as they react to different environmental impacts.

Leaving aside a question of how one can properly define a particular type of synchronization, we would like to consider two different views of the same kind of synchronized phenomena. One is the phase space view and the other is the time-series, observables-based view. Intermittent synchrony from the observational standpoint is the case, when the time-series of two oscillators appear to be "synchronized" (correlated in certain sense and in statistically significant manner) for some temporal episodes and non-synchronized for other temporal episodes. In the phase space view, intermittent synchrony may correspond to the case, where certain synchronized state (which may or may not be an invariant synchronization manifold) is not stable, but nevertheless trajectory enters a vicinity of this state relatively frequently and leaves it relatively slowly. We will discuss here both views and our recently developed approach for the analysis of this synchronization/desynchronization dynamics (see (6,7) for some original results).

A straightforward approach to the temporal variability of the synchrony strength is to use sliding short temporal windows for the analysis. Synchrony within a short window may be checked for statistical significance (8). Even though this approach provides some temporal resolution, this resolution is not expected to be very high because synchrony is not an instantaneous phenomenon. As the window size becomes shorter, the statistical significance of synchrony is harder to estimate. In the phase space, a trajectory may leave from and enter to a vicinity of a synchronized state in relatively short time-intervals, much shorter than those required for statistically significant estimation of synchrony.

Our new approach is focused on the variability on shorter time-scales. This is not, of course, a detection of "instantaneous synchrony", which does not exist. What we can do is to detect the presence of some synchrony (in some specific sense) on sufficiently long temporal interval, then to look at how close two states or two observables are at each instant of time and detect if they are in synch or not at any particular cycle of oscillations.

**Methodological considerations**

After confirming that some synchrony is indeed present (for example, one can estimate the phase locking index and confirm its significance with appropriate statistical test, like in (6)), one may inspect if two systems (signals) are in synchronous state or not at each instant of time. This can be done with various operational definitions of synchrony and using different definition of how close the signals or states should be to each other to be considered synchronous. This may depend on a particular system under study. As an example, we will consider a case of intermittent phase locking.

The phase can be extracted from the "good" oscillatory data (the data with relatively narrow and prominent peak in spectrum) in several ways and we use Hilbert phase (see (1)). Using Hilbert transform one obtains an analytic signal $\zeta(t)$ from real time series $x(t)$

$$\zeta(t) = x(t) + i\,\bar{x}(t)$$
$$\bar{x}(t) = H(x) = \frac{1}{\pi}\,\text{P.V.}\int_{-\infty}^{\infty}\frac{x(\tau)}{t-\tau}\,d\tau,$$

and the phase of the analytic signal $\zeta(t)$, say $\varphi(t)$, is the Hilbert phase of the time series. It is given by

$$z(t) = \frac{\zeta(t)}{\|\zeta(t)\|} = e^{i\varphi(t)}.$$

This phase is defined modulo ($2\pi$) here. If the phase difference between two oscillators tends to be close (in some specific sense) to some constant value, then we can consider this as a synchronized dynamics.

One can compute a fairly standard phase locking index for two phases $\varphi_1(t)$ and $\varphi_2(t)$

$$\gamma = \left\|\frac{1}{N}\sum_{j=1}^{N}e^{i\Phi_j}\right\|^2$$

where $\Phi_j = \varphi_1(t_j) - \varphi_2(t_j)$ and $N$ is the number of data points (for the case of discrete time-series). This index varies between zero (no phase locking) and one (perfect phase locking) and one can analyze whether it has statistically significant non-zero value (e.g., using appropriately generated surrogate data (8)). Note that while dealing with experimental data one might want to use some kind of signal-to-noise ratio criteria before extracting narrow-spectrum signal to confirm the presence of oscillations in the otherwise wide-spectrum processes.

To simplify further analysis we suggest considering return maps for the phase difference. In other words, we are considering if the phase difference is close to its preferred (locked) state or not once per cycle of oscillations. How close it should be depends on a particular problem under consideration. We will consider the case, where we require the phase difference to be within $\frac{\pi}{2}$ of the preferred phase difference.

More specifically, let $\varphi_1(t)$ and $\varphi_2(t)$ be phases of two signals. Set up a checking point for $\varphi_1(t)$, say $\varphi_0 = const$, and record the value of $\varphi_2(t)$ whenever $\varphi_1(t)$ crosses the level of $\varphi_0$ in positive direction. This yields a set of consecutive phase values $\{\phi_i\,|\,i = 1, 2, \ldots, M\}$. Then the set of phase differences between two oscillators is given by $\{\phi_i - \varphi_0|\,i = 1, 2, \ldots, M\}$. Without loss of generality, assume that $\varphi_0 = 0$. Then the return map is obtained by plotting $\phi_{i+1}$ vs $\phi_i$ for $i = 1, 2, \ldots, M - 1$ in ($\phi_i$, $\phi_{i+1}$) phase space (see Fig. 1). In two extreme cases, fully synchronized or fully desynchronized cases, the return map either would yield a single point on the diagonal $\phi_{i+1} = \phi_i$ or would fill the ($\phi_i$, $\phi_{i+1}$)- space. Since we are considering sufficiently strong synchrony, return map yields a cluster on the $\phi_{i+1} = \phi_i$ diagonal due to the presence of synchronous dynamics with some deviations from it. For the uniformity of the analysis, let us move the center of the cluster to a fixed position, a point with coordinates ($\frac{\pi}{2}, \frac{\pi}{2}$). The ($\phi_i$, $\phi_{i+1}$) phase space is a 2D torus and the shift of the cluster to a new center does not disrupt the intrinsic temporal structure of synchrony.

[Figure 1]

Next the phase space is partitioned into several regions. There may be different ways to partition it into synchronized and nonsynchronized regions. However, with our $\pm\pi/2$ brackets, it is reasonable to partition it into four equal regions (see Fig. 1). The first region (I) is considered to be a synchronized state while other regions (II, III, and IV) are considered as desynchronized states.

To study the dynamics of return maps, we define the transition rates $r_{1,2,3,4}$ for transitions between four regions of the map (7). The rate $r_1$ is the ratio of the number of trajectories escaping the region I toward the region II to the number of all points in the region I. Similarly, $r_2$ is the ratio of the number of trajectories escaping the region II toward the region IV to the number of all points in the region II; $r_3$ is the ratio of the number of trajectories escaping the region III toward the region IV to the number of all points in the region III; $r_4$ is the ratio of the number of trajectories escaping the region IV toward the region I to the number of all points in the region IV. Note that the transitions of return maps are from ($\phi_i$, $\phi_{i+1}$) to ($\phi_{i+1}$, $\phi_{i+2}$) (that is the second coordinate in the current state, $\phi_{i+1}$, is the first coordinate in the next state) so that certain transitions are not possible (see Fig. 1).

The transition rates vary between zero and one. If $r_1$ is 0, then the system is in a synchronized state. The higher $r_1$ is, the lower synchrony strength is. The rate $r_1$ is essentially an inverse of the mean duration of the synchronized episode (laminar interval, as it would be called in the language used to describe intermittency) if it is measured in the number of cycles of oscillations. Larger values of other three rates promote faster return to the synchronized state and thus shorter desynchronization episodes.

One can also compute the distribution of durations of desynchronization events (7). If time is measured in cycles of oscillations, duration can be defined as the number of cycles that the system spends away from the synchronized state (region I) minus one. The shortest desynchronization event corresponds to the shortest path II → IV → I. We will call this the desynchronization episode lasting one cycle of oscillations (in two cycles the oscillations are again close to the phase locked state).

If transitions are independent, then the distribution of the durations may be obtained from the transition rates. At least for some cases the analysis of experimental data suggests that transitions may be close to independent (6, 9). Thus one can apply Markov chain model. The transition matrix will have a form

$$\begin{pmatrix} (1-r_1) & r_1 & 0 & 0 \\ 0 & 0 & (1-r_2) & r_2 \\ 0 & 0 & (1-r_3) & r_3 \\ r_4 & (1-r_4) & 0 & 0 \end{pmatrix}$$

and the Markov chain formalism will provide statistical description of the synchronization/desynchronization dynamics (which may be especially useful if one deals with an ensemble of synchronized systems).

## An analysis of a simple model system

To illustrate some of the ideas discussed above, following (7) we will consider an example of a very simple coupled system: two coupled skewed tent maps. While this example may be ill-suited to study phase synchronization (10), it helps to illustrate the major ideas of our approach using the very simple system (piece-wise linear maps). Consider a skew tent map

$$f(a, x) = \begin{cases} \dfrac{x}{a}, & \text{if } 0 \leq x \leq a, \\ \dfrac{1-x}{1-a}, & \text{if } a < x \leq 1, \end{cases}$$

where $0 < a < 1$. We consider two such maps, described by variables $x$ and $y$, linearly coupled in the following way:

$$x(t+1) = (1-\varepsilon)f(a, x(t)) + \varepsilon f(a, y(t)),$$
$$y(t+1) = \varepsilon f(a, x(t)) + (1-\varepsilon)f(a, y(t)),$$

where $\varepsilon$ is the coupling strength. The difference of the variables of two maps $\Phi(t) = y(t) - x(t)$ may serve as a proxy for the phase difference. The synchronous state is $x = y$. It becomes stable for $\varepsilon$ larger than a critical value $\varepsilon_c$. Two Lyapunov exponents ($\lambda(a)$ and $\lambda_\perp(a, \varepsilon)$) can be computed analytically (1) and are not changed if $a$ is changed into $(1-a)$, i.e. they are symmetrical about $a = 1/2$.

[Figure 2]

Therefore two different pairs of maps with symmetrical values of *a* have the same values of Lyapunov exponents (in particular, the same value of $\lambda_\perp(a, \varepsilon)$, which characterizes the stability of the synchronous state). Thus they have the same expansive/contractive properties *on the average*. But the two systems are different. In one case, the map is strongly expansive in a small area of the phase space, while in the other case the map is less expansive, but the corresponding area is larger. As a result, while the properties of synchronized dynamics are the same, the properties of the desynchronized dynamics (such as values of the transition rates) are different between the two systems (7). The transition rates $r_{1,2,3,4}$ and the distributions of desynchronization episode durations are markedly different (Fig. 2).

**Applications**

A series of applications of these ideas have been published recently in the area of neuroscience. This is probably not very surprising. Neural synchrony is believed to be critical for a variety of cognitive and motor phenomena (3,11,12). Neural systems are very efficient to process signals from and react to quickly changing environment so that it may be natural for them to be in an intermittent state. Neural synchrony is rarely very strong for a prolonged interval of time. Moreover, excessive neural synchrony is associated with many brain disorders, such as Parkinson's disease or schizophrenia (6,11,13,14,15).

Studies of the mammalian brain in different conditions suggest that the synchronous neural activity is usually punctuated by numerous, but short desynchronization episodes. Arbitrary coupled oscillators do not necessarily exhibit this kind of desynchronization dynamics (7, also see Fig. 2). However, neural oscillators exhibit short desynchronization pattern in various settings. It was observed in electroencephalographic (EEG) recordings in healthy human subjects (9), in the extracellular recordings of neural spiking and local field potential (LFP) recordings from the subcortical areas in the brain of Parkinsonian patients (6,15), and in the LFP recordings from the cortex and hippocampus in brain of healthy and drug-addicted rodents (16).

Similarly, synchronization between cardiac and respiratory rhythms exhibits prevalence of short desynchronizations, although to a smaller degree than those observed in the neuronal systems. The latter analysis required the generalization of the approach to $1:m$ frequency locking cases as cardiorespiratory synchronization is rarely at 1:1 frequency ratio (17).

These findings may suggest that short desynchronization dynamics is universal in living neural networks and is likely to contribute to essential neural functions. One possibility is that short desynchronization dynamics may facilitate the formation and breaking of functional neural ensembles whenever needed at a small expense in a short time.

Some of these results of experimental data analysis (including short desynchronization dynamics) have been reproduced in modeling studies (18,19,20) and were used to compare the experimental and modeling dynamics (18,21). The information about desynchronizations allows matching the phase spaces of model and real systems away from the synchronized state (see Discussion below).

The analysis of the desynchronization episodes may be helpful to detect small changes in the coupled oscillatory systems. It was found that the ratio of the number of short desynchronization to long desynchronizations is sensitive to the early development of drug addiction, while average synchrony strength is not significantly affected by initial drug delivery (16).

Finally, it is interesting to note, that in the case of the EEG data from healthy individuals, the resulting transition rates are quite close to values, which satisfy the following condition: $(1 - r_1) = r_2 = r_3 = r_4$. If this condition is strictly satisfied, then the eigenvalues of the corresponding Markov chain transition matrix are 0 (multiplicity 3) and 1. Thus, if the stationary (and maybe optimal in some sense) state of pairs of synchronized oscillators is perturbed, there will be a very fast convergence back to this stationary state (9).

**Discussion**

The approach discussed here is not expected to be useful, if the coupled oscillators are in the completely synchronized state (or very close to this state). However, as we discussed above, in many cases the synchronization observed in nature is quite weak. In these cases, the oscillators spend substantial amount of time in the desynchronized state. To better understand the dynamics of this relatively weak synchrony, one needs to characterize not only the properties of synchronized state, but also properties of desynchronized states and transitions between them. This is what the discussed approach is aimed at.

Traditionally, the focus of synchronization analysis is on the stability of the synchronized state by using, for example, Lyapunov exponents. However, if the system is weakly synchronized and does not spend much time in the vicinity of the synchronized state, the utility of the knowledge about the synchronized state may be limited because the trajectory spends a lot of time on the periphery of the phase space. The dynamics in these parts of the phase space (i.e., during desynchronized episodes) needs to be described too.

Depending on how the synchronized state loses its stability, different types of intermittent synchronization are possible. But they are universal in a sense that a particular bifurcation of synchronized state may lead to a particular type of intermittency regardless of some features of the coupled oscillators. Unlike the universality of the synchronized episodes, the desynchronized episodes are not expected to be universal. The mechanism of reinjection of the trajectory back to the vicinity of the synchronized state will depend on the specific properties of oscillators and coupling. Different oscillatory systems with the same types of intermittency and the similar strength of phase locking may exhibit different temporal structures of synchronization/desynchronization events (7). These differences may be detected and described using the methods reviewed here.

In light of the apparent lack of universality of desynchronizations, the persistence of observations of short desynchronization events in the neural activity of the brain discussed in the previous section becomes very interesting. It points to the potential significance of short desynchronization dynamics. Neural networks of the brain may have evolved in such a way that the mechanisms of the reinjection to the vicinity of the synchronized state promote short desynchronization dynamics. This would complement recent observations and conjectured functional significance of high variability of synchrony in critical dynamics of the cortex of the brain (21,22).

Finally, we would like to note that the use of transition rates (or other related characteristics of the dynamics) may assist in the matching models to experimental data (as was done in (18,23)). Not only the average synchrony strength may be matched, but also the properties of desynchronizations may be matched too. This helps to match the structure of the areas of the phase space, which are away from the synchronized state, but where, nonetheless, the system spends substantial amount of time.

**Acknowledgments**
This study was supported by NIH grant R01NS067200 (NSF/NIH CRCNS program) and by Indiana University Collaborative Research Grant.

**Disclosures**
No conflicts of interest are declared by the authors.

**References**

1. Pikovsky A, Rosenblum M, Kurths J. *Synchronization: A Universal Concept in Nonlinear Sciences*. Cambridge, UK: Cambridge University Press (2001).
2. Glass L. Synchronization and rhythmic processes in physiology. *Nature* (2001) **410**: 277-284.
3. Buzsáki G, Draguhn A. Neuronal oscillations in cortical networks. *Science* (2004) **304**: 1926–1929.
4. Rabinovich MI, Varona P, Selverston AI, Abarbanel HDI. Dynamical principles in neuroscience. *Rev Mod Phys* (2006) **78**: 1213.
5. Nowotny T, Huerta R, Rabinovich MI. Neuronal synchrony: Peculiarity and generality. *Chaos* (2008) **18**: 037119.
6. Park C, Worth RM, Rubchinsky LL. Fine temporal structure of beta oscillations synchronization in subthalamic nucleus in Parkinson's disease. *J Neurophysiol* (2010) **103**: 2707-2716.
7. Ahn S, Park C, Rubchinsky LL. Detecting the temporal structure of intermittent phase locking. *Phys Rev E* (2011) **84**: 016201.
8. Hurtado JM, Rubchinsky LL, Sigvardt KA. Statistical method for detection of phase-locking episodes in neural oscillations. *J Neurophysiol* (2004) **91**: 1883-1898.
9. Ahn S, Rubchinsky LL. Short desynchronization episodes prevail in synchronous dynamics of human brain rhythms. *Chaos* (2013) **23**: 013138.
10. Rosenblum MG, Pikovsky AS, Kurths J. Comment on "Phase synchronization in discrete chaotic systems". *Phys Rev E* (2001) **63:** 058201.
11. Uhlhaas PJ, Singer W. Neural synchrony in brain disorders: Relevance for cognitive dysfunctions and pathophysiology. *Neuron* (2006) **52**: 155-168.
12. Fell J, Axmacher N. The role of phase synchronization in memory processes. *Nat Rev Neurosci* (2011) **12**: 105-118.
13. Schnitzler A, Gross J. Normal and pathological oscillatory communication in the brain. *Nat Rev Neurosci* (2005) **6**: 285–296.
14. Uhlhaas PJ, Singer W. Abnormal neural oscillations and synchrony in schizophrenia. *Nat Rev Neurosci* (2010) **11**: 100–113.
15. Rubchinsky LL, Park C, Worth RM. Intermittent neural synchronization in Parkinson's disease. *Nonlinear Dyn* (2012) **68**: 329-346.

16. Ahn S, Rubchinsky LL, Lapish CC. Dynamical reorganization of synchronous activity patterns in prefrontal cortex–hippocampus networks during behavioral sensitization. *Cereb Cortex* (2014), accepted. doi: 10.1093/cercor/bht110
17. Ahn S, Solfest J, Rubchinsky LL. Fine temporal structure of cardiorespiratory synchronization. *Am J Physiol Heart Circ Physiol* (2014) **306**: H755-H763.
18. Park C, Worth RM, Rubchinsky LL. Neural dynamics in parkinsonian brain: the boundary between synchronized and nonsynchronized dynamics. *Phys Rev E* (2011) **83:** 042901.
19. Park C, Rubchinsky LL. Intermittent synchronization in a network of bursting neurons. *Chaos* (2011) **21:** 033125.
20. Park C, Rubchinsky LL. Potential mechanisms for imperfect synchronization in parkinsonian basal ganglia. *PLoS ONE* (2012) **7**(12): e51530.
21. Yang H, Shew WL, Roy R, Plenz D. Maximal variability of phase synchrony in cortical networks with neuronal avalanches. *J Neurosci* (2012) **32**:1061-1072.
22. Meisel C, Olbrich E, Shriki O, Achermann P. Fading signatures of critical brain dynamics during sustained wakefulness in humans. *J Neurosci* (2013) **33**:17363-17372.
23. Dovzhenok A, Park C, Worth RM, Rubchinsky LL. Failure of delayed feedback deep brain stimulation for intermittent pathological synchronization in Parkinson's disease. *PLoS ONE* (2013) **8**(3): e58264.

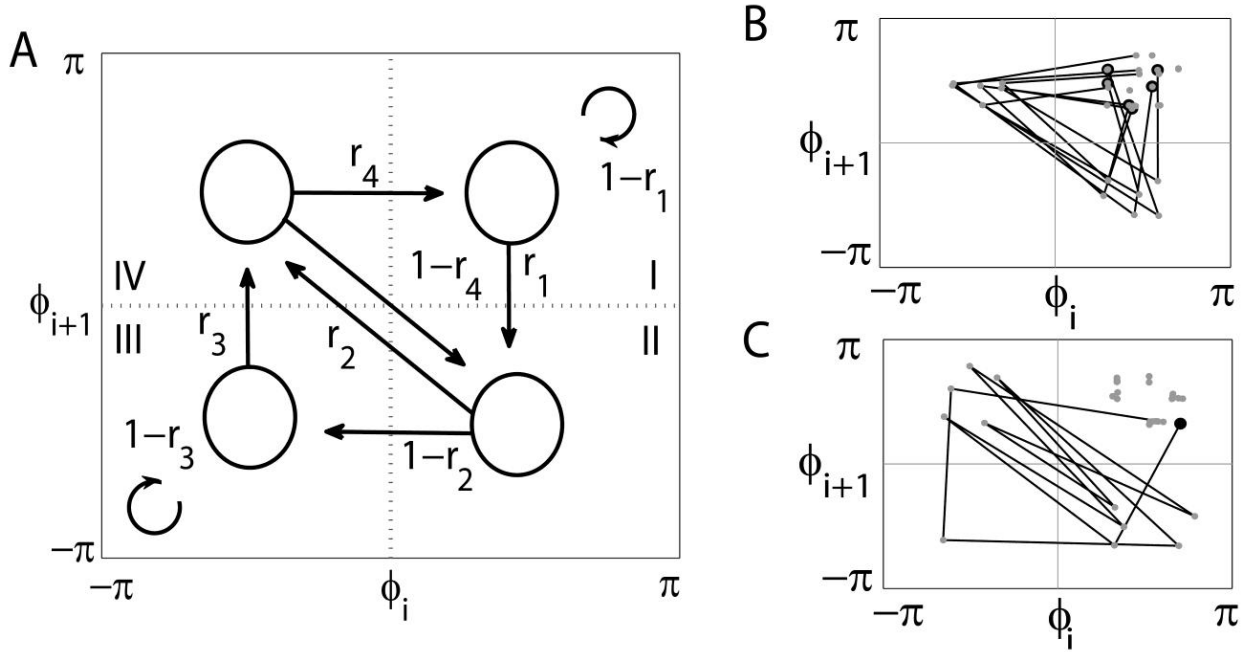

**Figure 1.** (A) Diagram of the $(\phi_i, \phi_{i+1})$ first-return map. The arrows indicate all possible transitions from one region to another and $r_{1,2,3,4}$ indicate the corresponding transition rates. After the uniform phase shift for all the data, the synchronized state is in the center of the region I (which is called synchronized state too) and three other regions are desynchronized states. (B) and (C) present examples of two extreme cases of dynamics. (B) presents numerous short desynchronizations. Whenever trajectory leaves synchronized region I, it follows the path II-IV-I and returns back to the synchronized state in the shortest possible way. Thus we have many short desynchronization episodes. (C) presents an opposite example of one, but very long desynchronization event. The average synchrony levels in both examples are very similar, but the temporal patterning of synchronization is very much different.

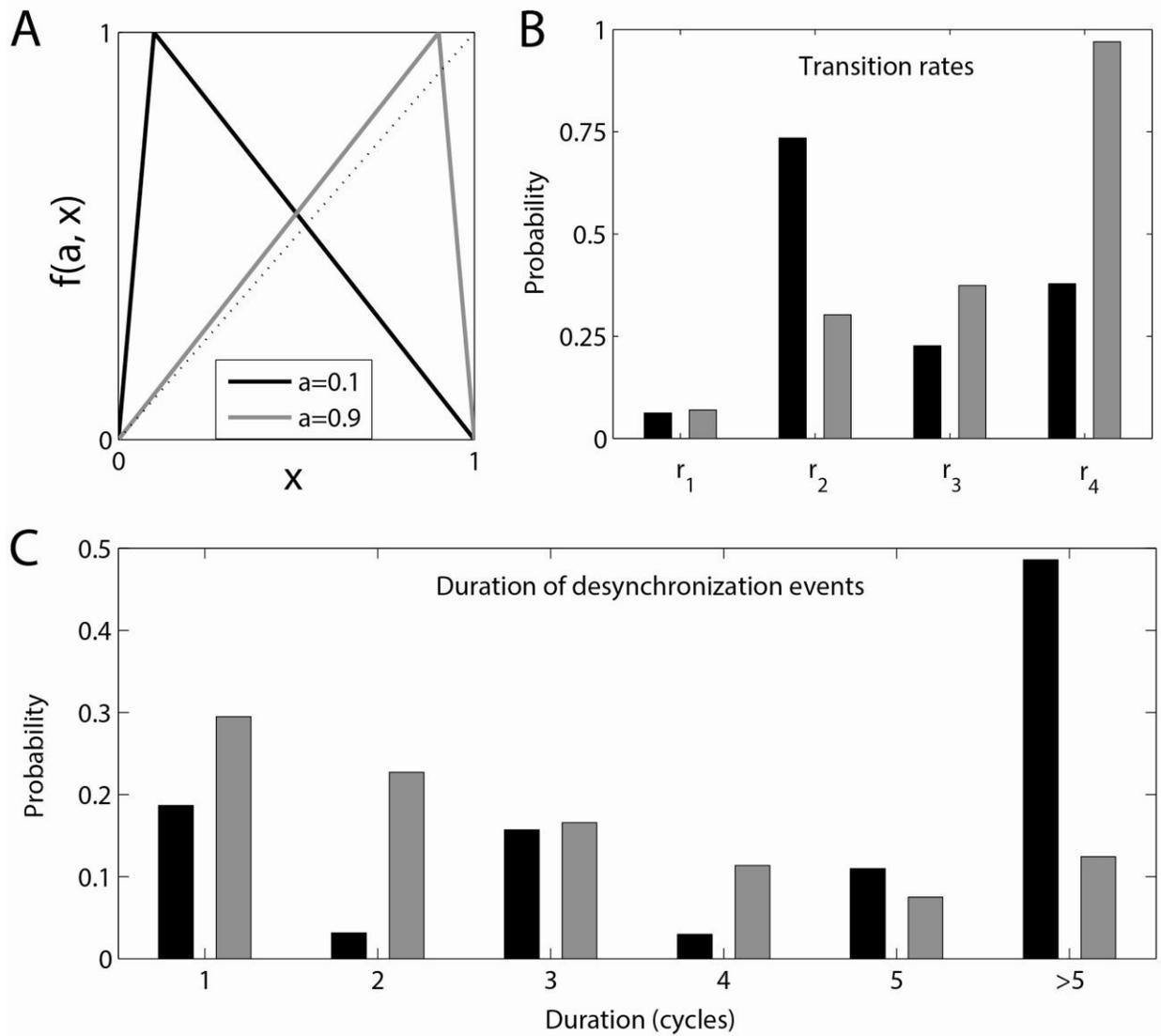

**Figure 2.** Two pairs of coupled skew tent maps with $a = 0.1$ (black) and $a = 1 - 0.1 = 0.9$ (gray). The coupling value $\varepsilon = 0.02$ and is less than $\varepsilon_c$. The coupled "black" and "gray" maps have identical Lyapunov exponents but different expansive/contractive properties in different areas of the phase space. (A) the maps, (B) transition rates, and (C) distribution of durations desynchronization events.